\documentclass[twocolumn,aps,epsfig,showpacs,floatfix]{revtex4}
\usepackage{epsfig}
\usepackage{dcolumn}
\usepackage{bm}

\begin{document}

\title{Consequences of energy conservation in relativistic heavy-ion collisions}

\author{B.B.Back} 
\affiliation{Physics Division, Argonne National Laboratory, Argonne, IL 60439, USA}

\begin{abstract}

Complete characterization of particle production and emission in relativistic heavy-ion collisions is in general not feasible experimentally. This work demonstrates, however, that the availability of essentially complete pseudorapidity distributions for charged particles  allows for a reliable estimate of the average transverse momenta and energy of emitted particles by requiring energy conservation in the process. The results of such an analysis for Au+Au collisions at $\sqrt{s_{\it NN}}$= 130 and 200 GeV are compared with measurements of mean-$p_T$ and mean-$E_T$ in regions where such measurements are available. The mean-$p_T$ dependence on pseudorapidity for Au+Au collisions at 130 and 200 GeV is given for different collision centralities. 
\end{abstract}

\vspace{0.5cm}

\maketitle 
            
\section{Introduction}

In recent studies of ultra-relativistic heavy-ion collisions at the Relativistic Heavy-Ion Collider (RHIC) at Brookhaven National Laboratory much attention has been devoted to measurements of particles and radiation emitted in the transverse direction. This focus is well justified because particles emitted in this direction are expected~\cite{Bjorken} to carry information about the region of highest energy density formed in the process, which is believed to consist of de-confined quarks and gluons~\cite{QGP}. Experimental evidence based on Au+Au collisions~\cite{BRAHMS_WP,PHOBOS_WP,STAR_WP,PHENIX_WP} indicates that a phase of matter with extreme energy density is indeed formed in such collisions, the properties of which correspond to an ideal fluid of quarks and gluons. 

The present paper is instead focussed on following the total energy in the collision and how it is distributed in the exit channel. It is interesting to note that the number of particles emitted at mid-rapidity increases only slowly as a function of collision energy~\cite{PHOBOS_PRL88}. The increase in collision energy by more than a factor of one hundred from the CERN SPS facility to RHIC thus result in only doubling of the mid-rapidity particle multiplicity. The bulk of the additional energy is carried off by particles emitted away from this region. Thus, in order to study the energy balance in the reaction it is necessary to measure particles emitted over essentially the full solid angle. The PHOBOS experiment at RHIC~\cite{PHOBOS_NIM} is unique in being close to achieve such coverage for charged particles and nearly complete distributions of charged particles in pseudorapidity space, $dN_{\it ch}/d\eta$, have been reported~\cite{PHOBOS_limfrag}. In this work, these and other data are used to show that energy conservation places severe constraints on the transverse momentum distributions away from midrapidity, outside the range where they can be measured. The results are used to calculate the pseudorapidity dependence of the mean transverse momentum, mean transverse energy and an estimate of the initial energy density in the spatial region from which particles with large longitudinal momenta originate.

\section{Entrance channel}
At relativistic energies, it is normally assumed that only a fraction of nucleons, $N_{\it part}$, of the incoming heavy ions take part in the collision. The non-participaring nucleons, the spectators, are assumed to continue unaffected by the collision between the participants and do therefore not contribute to the particle production. For the purpose of the present work, the energy of the spectators will therefore not be included in the energy accounting. At the 130 GeV and 200 GeV collision energies studied here, it is believed that the pseudorapidity distribution of the spectators falls outside the PHOBOS acceptance. On the other hand, it is assumed that all of the energy carried by the participants will appear in the exit channel. In a recent analysis of net baryons in the rapidity range $y$=0-3.2, the BRAHMS collaboration~\cite{BRAHMS_protons} have concluded that they retain $\sim$27\% of their initial energy, which might lead one to the conclusion that only $\sim$ 73\% should appear in the exit channel. It is important, however, to realize that also the partially stopped net baryons are included in the measured (and extrapolated) $dN/d\eta$ distribution of charged particles considered in this work as demonstrated in Sect. VII.  Consequently, one needs to account for all of the energy of the participants, namely
\begin{equation}
E_{\it in}=\sqrt{s_{\it NN}}\times N_{\it part}/2,
\end{equation}
where $\sqrt{s_{\it NN}}$ is the center-of-mass energy per nucleon pair and $N_{\it part}/2$ is the number of such pairs. The value of  $N_{\it part}$ is normally calculated in the Glauber model \cite{Glauber} for collisions at different impact parameters using experimental nucleon-nucleon cross sections. The centrality of heavy-ion collisions is typically given by the fraction of the total cross sections such that the 0-6\% centrality bin corresponds to essentially head-on collisions exhausting 6\% of the total cross section, the 6-15\% centrality bin is associated with the 9\% of the cross section with increasing impact parameter, etc. 

\section{Exit channel}

Since the analysis is based on measurements of pseudorapidity distributions of charged particles, $dN_{\it ch}/d\eta$, one should expect to account for only this fraction of the total energy, {\it i.e.}
\begin{equation}
E_{\it ch}^{\it out}=f_{\it ch}\times \sqrt{s_{\it NN}}\times N_{\it part}/2,
\end{equation}
where 
$f_{\it ch}$ is the fraction of exit particles that are charged.

The energy of a particle is $E=m_T\cosh y$, where $m_T=\sqrt{m^2+p_T^2}$ is the transverse mass, $m$ is the rest mass, $p_T$ is the transverse momentum, and $y$ is the rapidity of the particle. Using the identity $m_T \sinh y = p_T \sinh \eta$ one finds
\begin{equation}
E=m_T \cosh y = \sqrt{m^2+p_T^2\cosh^2 \eta}.
\end{equation}

The total energy of charged particles may be evaluated by summing over pions, kaons, and protons and integrating over the distribution in $p_T$ and $\eta$

\begin{equation}
E_{ch}^{out}=\int_{-\infty}^\infty \frac{dE}{d\eta} d\eta,
\end{equation}
where
\begin{equation}
\frac{dE}{d\eta}=\sum_{i=\pi,K,p} P_i\int_0^\infty \sqrt{m_i^2+p_T^2 \cosh^2\eta}\frac{d^2N_i}{dp_Td\eta}dp_T,
\end{equation}
and $P_i$ is the relative abundance of particles with mass $m_i$.

In the Appendix, the effects of computing $dE/d\eta$ (Eq. 5) under different assumptions are explored and it is shown that using an average mass $\langle m \rangle = \sum_{i=\pi, K,p} P_i m_i$ is an excellent approximation to an accurate evaluation of the individual contributions from the three particle species with different $\langle p_T \rangle$ values. The dependence of $\langle p_T \rangle$ on particle mass has been observed experimentally and is normally interpreted as a consequence of radial flow of the hot fireball. In the following, the expression
\begin{equation}
\frac{dE}{d\eta}=\frac{dN}{d\eta} \int_0^\infty \sqrt{\langle m \rangle ^2+p_T^2 \cosh^2\eta} \frac{dN}{dp_T} dp_T.
\label{eq9}
\end{equation}
is therefore used to find those $\eta$-dependences of $\langle p_T \rangle$ that satisfy energy conservation.

\section{Input data}

The $dN_{\it ch}/d\eta$ distributions the and corresponding values of $N_{\it part}$ were taken from Ref.~\cite{PHOBOS_limfrag}. The data and error bands were fitted to a high accuracy using three Gaussians, one centered at $\eta =0$, and two of identical shape centered at $\pm\eta_0$ using five fit parameters. 

\subsection{130 GeV}

Measurements by STAR~\cite{STAR_pt0,STAR_pt,STAR_0311017} have shown that the overall charged particle $p_T$-distributions near mid-rapidity follow a power-law behavior, 
\begin{equation}
\frac{dN}{dp_T}= \frac{p_T(n-2)(n-1)}{p_0^2}\left(1+\frac{p_T}{p_0}\right)^{-n},
\end{equation} 
where $n$ and $p_0$ are adjustable parameters. The values of $n$ and $\langle p_T \rangle=2p_0/(n-3)$ were taken from~\cite{STAR_0311017} and interpolated to match the centrality bins of the PHOBOS $dN/d\eta$ data.
In the Appendix it is shown that the total energy in the exit channel is very insensitive to the particle rest mass. This power-law shape of the $p_T$ spectra (in $\eta$-space) is therefore used for all three particle types (pions, kaons, protons/anti-protons).

The charged particle fraction was estimated using identified particle multiplicities $dN/dy|_{y=0}$ from Refs.~\cite{STAR_0311017} (pions), \cite{STAR_kaons} (kaons), and \cite{STAR_ppbar} (protons/anti-protons) and converting to pseudorapidity space $dN/d\eta|_{\eta=0}$ by using the fitted spectral shapes. The results were again interpolated to match the PHOBOS centrality bins.

\subsection{200 GeV}
 As yet, no analysis of charged hadron $p_T$-spectra has been published for 200 GeV Au+Au collisions. In this case it is therefore assumed that the centrality dependence of the $n$  parameter is identical to that found at 130 GeV. In order to estimate $\langle p_T \rangle_{\it \eta=0}$ and $f_{\it ch}$ for unidentified hadrons at this energy, the values of $dN/dy|_{y=0}$ of charged particles were taken from STAR~\cite{STAR_PRL92} and PHENIX~\cite{PHENIX_PRC69}, converted to pseudorapidity space using the fitted spectral shapes, interpolated to match PHOBOS centrality bins and averaged (with equal weights). Note that the transformation from rapidity to pseudorapidity space increases the values of $\langle p_T \rangle_{\eta=0}$ to be larger than those obtained directly from the $dN/dp_T$-spectra measured for a fixed rapidity interval.

The charged particle fractions, $f_{\it ch}$, were obtained from the same data according to the method used for 130 GeV. All resulting values are listed in Table I.

\section{Average transverse momentum}

In this section it is shown that the requirement of energy conservation and the fact that $\langle p_T \rangle$ is experimentally known at $\eta=0$ imposes severe restrictions on the $p_T$-distribution of particles emitted at large pseudorapidities. 

\begin{figure}
\epsfig{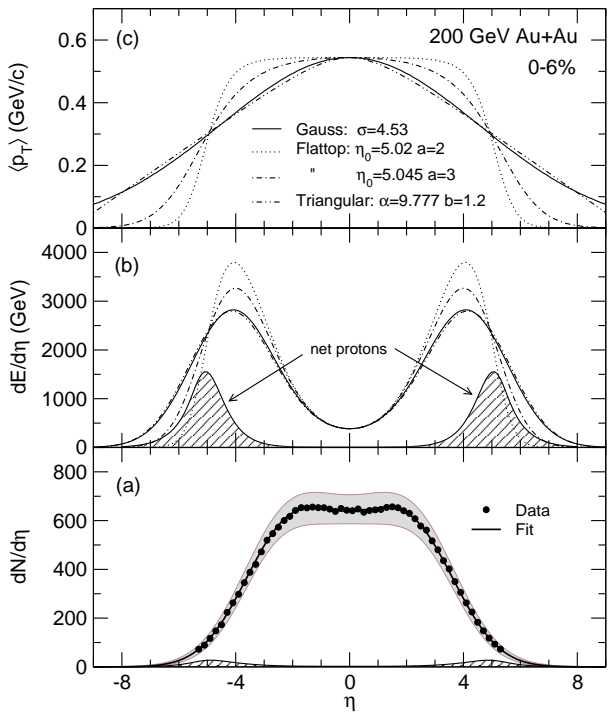}
\vspace{2mm}
\caption{Panel a) Pseudorapidity distrbution for 0-6\% central Au+Au collisions at 200 GeV~\cite{PHOBOS_limfrag} (solid points and error band). Panel b):  $dE/d\eta$ is plotted for different assumptions of the $\eta$ dependence on $\langle p_T \rangle$, which are shown in Panel c). The estimates of the $\eta$-distribution of net-protons and their energy derived in Sect. VII is shown a grey shading in panels b) and c), respectively.  }
\label{Fig1}
\end{figure}
 
First, it is shown that a constant (as a function of $\eta$) $\langle p_T \rangle$ does not conserve energy. By assuming that particles emitted at all $\eta$ values are characterized by $n$= 21.9 and $\langle p_T \rangle$ = 0.544 (GeV/c), corresponding to those measured at mid-rapidity for 200 GeV Au+Au collisions at 0-6\% centrality, Eq. 6 gives a total energy of $E_{ch}^{tot}$= 35,523 GeV, which is substantially larger than the $E_{ch}^{out}$ = 21,844 GeV expected in outgoing charged particles on the basis of energy conservation. One must therefore conclude that $\langle p_T \rangle(\eta)$ assumes smaller values (softer $p_T$-spectra) away from mid-rapidity. In the following, different shapes of the $\langle p_T \rangle$-dependence on $\eta$ are explored. Guided by predictions of theoretical models, such as HIJING~\cite{HIJING}, RQMD~\cite{RQMD}, and AMPT~\cite{AMPT} it is assumed that $\langle p_T \rangle$ falls off monotonically away from $\eta$=0. Only analytical functions which have this property are therefore considered.

Assuming a Gaussian $\eta$ dependence of $\langle p_T \rangle$ and requiring energy conservation allows for a determination of the standard deviation $\sigma$ of the Gaussian dependence. The solid curves in Fig. 1 represent a calculation of $dE/d\eta$ for 200 GeV central (0-6\%) Au+Au  assuming 
\begin{equation}
\langle p_T \rangle _\eta = \langle p_T\rangle_0 \exp(-\eta^2/2\sigma^2)
\label{gauss}
\end{equation}
using the parameters $\langle p_T\rangle_0 $ = 0.544 GeV/c,  $\sigma$=4.53. These parameters result in a total charged particle energy which reproduces the expected value of $E_{ch}^{out}$ = 21,844 GeV (obtained from Eq. 2 using parameters listed in Table I).

Although a Gaussian $\eta$-dependence of the mean transverse momentum probably approximates the actual dependence quite well other dependencies have also been studied. One parametrization is given by
\begin{equation}
\langle p_T \rangle _\eta= \langle p_T \rangle_0 \frac{1+\exp(-\eta_0^2 / a^2)}{1+ \exp[(\eta-\eta_0)(\eta+\eta_0)/a^2]}
\label{flattop}
\end{equation}
{\it i.e.} a flat-top distribution which falls off at $\pm \eta_0$ over a range of $a$.
 Since this form contains two adjustable parameters, a family of curves exists that will obey energy conservation as illustrated in Fig. 1 where such curves are shown for $a=2$ (dashed) and  $a=3$ (dotted-dashed) for the 200 GeV 0-6\% central Au+Au collision data. 

Finally, a triangular" shape given by

\begin{equation}
\langle p_T \rangle_\eta = \langle p_T \rangle_0 \frac{1-|\eta|/\alpha-b\exp(-|\eta|/b)/\alpha}{1-b/\alpha},
\end{equation}
where $\alpha$ controls the slope away from $\eta=0$ and $b$ the radius of curvature at midrapidity is explored. This function does not deviate significantly from the Gaussian when both are required to yield the expected total energy.
\begin{figure}[hbt] 
\epsfig{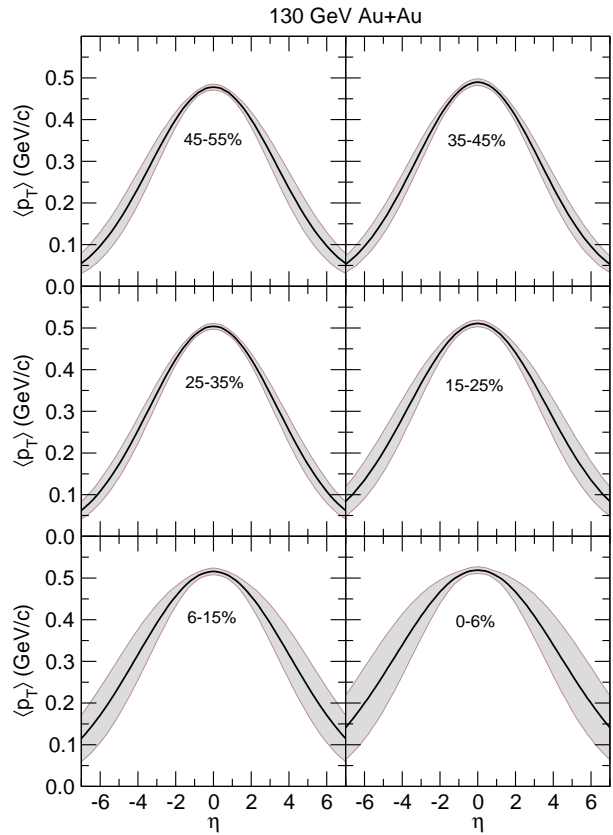}
\vspace{2mm}
\caption{Mean transverse momentum is shown as a function of pseudorapidity for eleven different centrality bins in Au+Au collisions at 130 GeV. Solid curves are Gaussian pseudorapidity depencences whereas dotted curves represent flattop dependencies, which also conserve energy during the collision.}
\label{TOF_eloss}
\end{figure}

It is interesting to note that the four curves shown in Fig. 1c representing different $\eta$ dependences of $\langle p_T \rangle$, all converge at at a value of $\langle p_T \rangle$ = 0.312 GeV/c at $\eta$ = 5.05. This value of $\langle p_T \rangle$ value may be considered a direct experimental measurement because it is independent of the assumed functional form of the $\langle p_T \rangle$-dependence on $\eta$.

\begin{figure}[hbt]
\epsfig{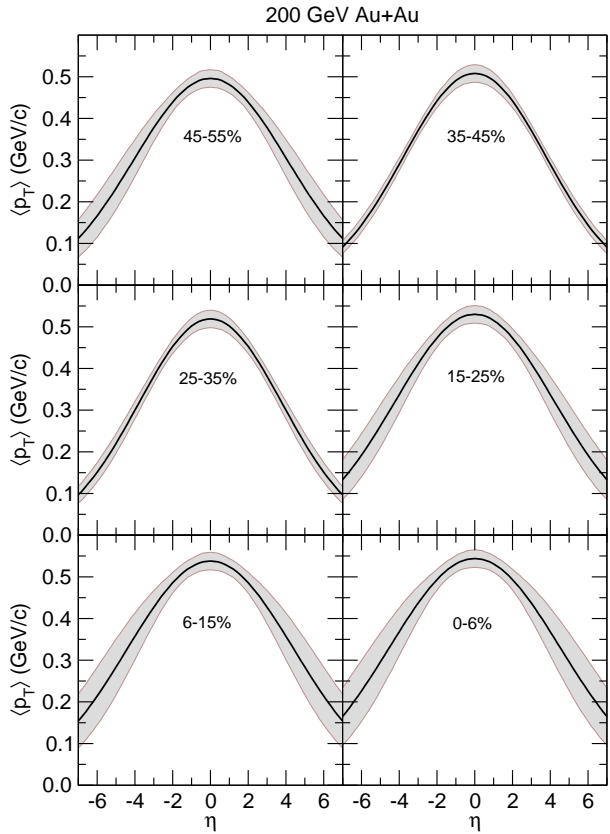}
\vspace{2mm}
\caption{Same as Fig. 2, but for 200 GeV.}
\label{TOF_eloss}
\end{figure}

\begin{table*}[tb]  
\caption{Summary of relevant parameters used in the analysis are listed as a function of the collision centrality. The origin of the input data ($N_{\it part}, f_{\it ch}, n, \rm{and} \langle p_T \rangle_{\eta=0}$) is detailed in the text. The last column list the $\sigma$ of the Gaussian $\langle p_T \rangle (\eta)$ curve (Eq.~\ref{gauss}) that are consistent with the value of $E_{\it tot}$. 
}
\vspace{3mm}
\begin{ruledtabular}
\begin{tabular}{cccccc|c}
$\sqrt{s_{\it NN}}$=130 GeV&&&&&&Gaussian\\
\hline
  Bin & $N_{\it part}$& $f_{\it ch}$ & $n$& $E_{\it ch}^{\it out}$&  $\langle p_T \rangle_{\eta=0}$&$\sigma$\\
	& & & & (GeV) & (GeV/c) & \\
\hline
0-6\%   & 340$\pm$11 &0.627$\pm$0.017& 22.1&13857$\pm$585 & 0.519$\pm$0.008&4.34\\
6-15\%  & 275$\pm$9  &0.627$\pm$0.017& 20.7&11208$\pm$476 & 0.516$\pm$0.008&4.04\\
15-25\% & 196$\pm$8  &0.627$\pm$0.020& 18.5& 7988$\pm$414 & 0.511$\pm$0.008&3.68\\
25-35\% & 136$\pm$6  &0.625$\pm$0.023& 17.2& 5525$\pm$317 & 0.504$\pm$0.007&3.42\\
35-44\% &  90$\pm$5  &0.624$\pm$0.029& 16.4& 3650$\pm$264 & 0.490$\pm$0.008&3.32\\
45-55\% &  59$\pm$5  &0.622$\pm$0.034& 14.8& 2385$\pm$208 & 0.478$\pm$0.007&3.36\\
\hline
$\sqrt{s_{\it NN}}$=200 GeV&&&&&&Gaussian\\
\hline
  Bin & $N_{\it part}$& $f_{\it ch}$ & $n$&$E_{\it ch}^{\it out}$&  $\langle p_T \rangle_{\eta=0}$&$\sigma$\\
	& & & &(GeV) & (GeV/c) & \\
\hline
0-6\%   & 344$\pm$11 &0.635$\pm$0.025& 22.1&21844$\pm$1108 & 0.544$\pm$0.021&4.53\\
6-15\%  & 274$\pm$9  &0.636$\pm$0.026& 20.7&17426$\pm$912  & 0.538$\pm$0.021&4.42\\
15-25\% & 200$\pm$8  &0.636$\pm$0.027& 18.5&12720$\pm$742  & 0.530$\pm$0.021&4.21\\
25-35\% & 138$\pm$6  &0.638$\pm$0.029& 17.2& 8804$\pm$554  & 0.519$\pm$0.021&3.82\\
35-45\% &  93$\pm$5  &0.639$\pm$0.030& 16.4& 5943$\pm$424  & 0.508$\pm$0.021&3.78\\
45-55\% &  59$\pm$5  &0.640$\pm$0.031& 14.8& 3776$\pm$315  & 0.496$\pm$0.021&4.06\\
\end{tabular}
\label{table1}
\end{ruledtabular}
\end{table*}

\section{Results}

The Gaussian $\langle p_T \rangle(\eta)$ functions, consistent with energy conservation, are shown for 130 and 200 GeV Au+Au collisions in Figs. 2 and 3 for six different centrality bins in each case. The parameters describing these functions are listed in Table I. Based on $p_T$ spectra, measured by BRAHMS \cite{BRAHMS_mesons,BRAHMS_protons} over a limited (pseudo)-rapidity range it is believed that the Gaussian shape is realistic. 

The error bands were obtained by propagating the errors on the experimental input parameters used in the analysis. These are: $\delta \langle p_T \rangle, \delta E_{\it ch}=\sqrt{(\delta f_{\it ch})^2+(\delta N_{\it part})^2}$ and $\delta N_{\it ch}$. The uncertainty on the mean $p_T$ of hadrons, $\delta \langle p_T \rangle$, was obtained from Refs. \cite{STAR_0311017,PHENIX_PRC69,STAR_PRL92}. The uncertainty on the total charged particle energy in the exit channel, $\delta E_{\it ch}$, has two sources, namely the uncertainty on the charged-particle fraction, $\delta f_{ch}$, which has been propagated from the measured values of $dN/dy$ for identified $p_T$ spectra of pions, kaons and protons and converted to the pseudorapidity space using the relevant $p_T$ fit functions. The second term, $\delta N_{\it part}$, was taken directly from Ref.~\cite{PHOBOS_limfrag}, as was the last component $\delta N_{\it ch}$. The final error on the $\langle p_T \rangle(\eta)$ functions were computed as a function of pseudorapidity. Clearly, the contribution from  $\delta \langle p_T \rangle$ dominates near $\eta=0$, whereas the errors on $dN_{\it ch}/d\eta$ and $E_{\it ch}$ prevail at larger values of $|\eta|$. In comparison to the uncertainty on experimental input data, the errors stemming from using an average rest mass of all hadrons ($m_0\sim 0.208$ GeV/c$^2$) are insignificant as demonstrated in the Appendix.

\begin{figure}[hbt]
\epsfig{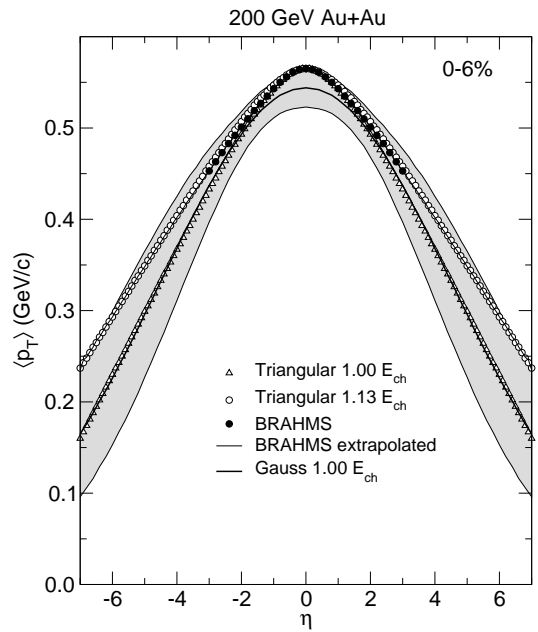}
\vspace{2mm}
\caption{Comparison of mean-$p_T$ obtained from different methods. See text for details.}
\label{TOF_eloss}
\end{figure}

\section{Comparison to BRAHMS results}

In recent publications, the BRAHMS collaboration has presented transverse momentum spectra and rapidity densities for pions, kaons~\cite{BRAHMS_mesons} and protons/anti-protons\cite{BRAHMS_protons} for rapidities ranging from $y=0$ to $y\sim3.4$. It is therefore of interest to compare these data to the results of the present analysis. Using double-gaussian fits to the BRAHMS data on $\langle p_T \rangle$ and $dN/dy$, and by using the spectral functions (exponential for pions and kaons and Gaussian for protons/anti-protons) for the shape of the $p_T$-spectra, the values of $\langle p_T \rangle (\eta)$ has been calculated over the full pseudo-rapidity range. The results are shown as solid circles in Fig. 4 in the range where experimental data are available and as a solid curve in the extrapolated region. The BRAHMS data fall within the error band of the present analysis, but lie in the upper range near mid-rapidity. The open circles represent a triangular ($\alpha$=11.2, b=1.2) $\eta$-dependence, which exceeds energy conservation by 13\%, but follows the extrapolated BRAHMS data quite closely. The open triangles achieve energy conservation and are seen to follow the Gaussian shape obtained in the analysis presented in the previous section (note that the slightly higher value at $\eta$=0 accounts for an insignificant portion of the total energy balance).

In a recent analysis, the BRAHMS collaboration has found that $\sim27\pm 5\%$ of the initial energy remains in net baryons after the collision\cite{BRAHMS_protons}. In agreement with predictions by the HIJING/B$\overline{\rm B}$ model ~\cite{HIJING_BB2} it is found that the net-protons are emitted in the region $y\sim 4\pm 2$, which explains why a relatively small number of particles can account for such a large fraction of the available energy. A simple model which approximates the rapidity distribution of net-baryon predicted by Ref. \cite{HIJING_BB2} as 
\begin{eqnarray}
\nonumber \frac{dN}{dy}&=&2.2\cosh y[e^{-(\sinh y-26)^2/1800}+e^{-(\sinh y+26)^2/1800}]\\
&&+6 e^{-y^2/2}
\end{eqnarray}
and uses an exponential $p_T-$distribution, $dN/dp_T=p_T e^{-p_T/t}$ with $t=0.5 e^{-y^2/72}$ (MeV/c) also accounts well for the predicted \cite{HIJING_BB2} rapidity distribution of net-baryon energy per participant, $dE/dy$, see Fig. 5. This model also predicts the pseudorapidity distribution of net baryons and their energy shown as dashed curves in Fig. 5. The pseudo-rapidity distributions of net protons and the energy they carry according to this simple model are also shown as shaded regions in Figs. 1a and 1b, respectively. It is evident that the predicted net proton distribution is easily accomodated within the measured $dN_{\it ch}/d\eta$ distribution. The associated energy appears, however, to exclude a range of $\langle p_T \rangle$ dependencies using the flattop function with $a < 3$. Functions with a sharper maximum at $\eta=0$, such as the Gaussian or triangular functions studied here, are clearly preferred.

\begin{figure}[hbt]
\epsfig{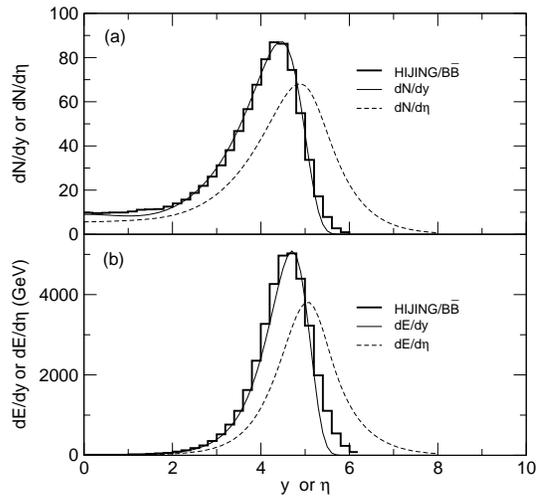}
\vspace{2mm}
\caption{Panel a): rapidity distributions of net baryons are shown for HIJING/B$\overline{\rm B}$ (solid histogram) and an analytical fit using Eq. 11 (solid curve). The dashed curve represents the pseudorapidity distribution corresponding to the fit. Panel b): The corresponding energy distributions are shown. }
\label{fig5}
\end{figure}

\section{Comparison to theory}
The results concerning the $\langle p_T \rangle$ dependence on pseudorapidity obtained in the previous section were based almost exclusively on constraints of experimental data; the measured $dN_{\it ch}/d\eta$  and net proton distributions in pseudo-rapidity space. It is, however, of interest to compare these results to predictions of theoretical models of particle production in relativistic heavy-ion collisions. In a recent work, Topor-Pop {\it et al.}~\cite{HIJING_BB1} have presented the predictions for the transverse energy of the HIJING/B$\overline{\rm B}$ and RQMD~\cite{RQMD} models. 

\begin{figure}[hbt]
\epsfig{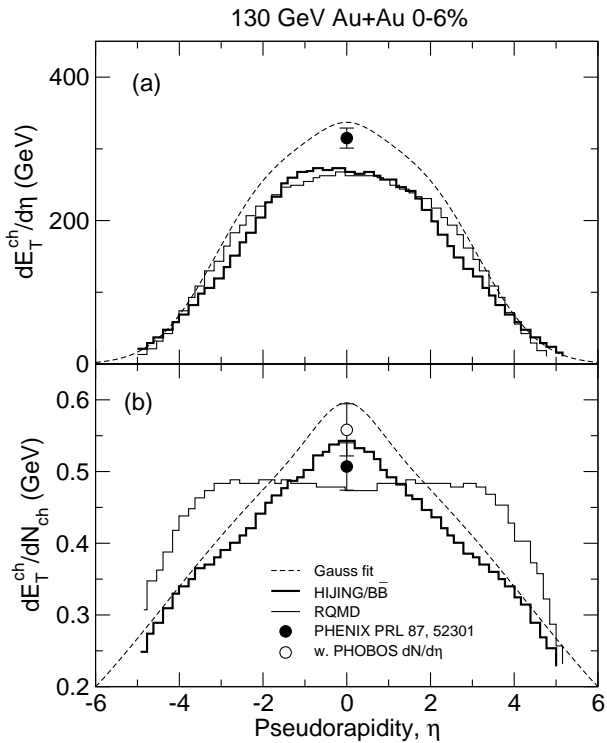}
\vspace{2mm}
\caption{Comparison of charged-particle transverse energy with predictions from the HIJING/BB model. See text for details.}
\label{TOF_eloss}
\end{figure}

The mean transverse energy of charged particles as a function of pseudorapidity obtained in the present analysis using a Gaussian $\eta$-dependence of $\langle p_T \rangle$ (dashed curve) is in Fig. 6a compared to those predicted by the HIJING/B$\overline{\rm B}$ (thick histogram) and RQMD (thin histogram) models. The overall distribution of transverse energy obtained in the present analysis agrees reasonably well with the model predictions. The value at mid-rapidity exceeds the prediction somewhat, which is also in agreement with the measurement of Ref. \cite{PHENIX_Et}. The latter has been multiplied by a factor of $f_{\it ch}$=0.627 to account for the fact that only charged particles are considered here; the PHENIX experiment refers to both charged and neutral particles. Fig. 6b shows a comparison of the charged particle transverse energy per charged particle, {\it i.e.} $(dE^{\it ch}_T/d\eta)/(dN_{\it ch}/d\eta)$ between the present analysis and the HIJING/B$\overline{\rm B}$ and RQMD predictions. While the RQMD model predicts a rather flat dependence on pseudorapidity, we find that the nearly triangular shape derived from the present analysis is quite well reproduced in the HIJING/B$\overline{\rm B}$ calculations. The closed and open circles in Fig. 6b represent values obtained using the PHENIX~\cite{PHENIX_dndeta} and PHOBOS~\cite{PHOBOS_limfrag} data on $dN_{\it ch}/d\eta$, respectively. The latter is in better agreement with the present analysis (dashed curve), which is based on the experimentally derived value of $\langle p_T \rangle_{\eta=0}$ listed in Table I.

\section{Energy density}

A crucial prerequisite for achieving the quark-gluon plasma phase is the initial energy density 
produced in the collision. Based on lattice QCD calculations \cite{Karsch} it is estimated that the transition to this phase occurs at an energy density of $\epsilon_{tr}\sim$ 0.7-1.0 (GeV/fm$^3$). Different estimates of the maximum energy density reached in central Au+Au collisions at RHIC energies \cite{PHENIX_Et,PHOBOS_WP} all indicate that the predicted threshold value of $\epsilon_{tr}$ in these collisions is exceeded by a large factor dependent on the assumed value of the equilibration time $\tau_0$, for which a conservative value of $\tau_0$ = 1 (fm/c) is often used. Estimates of the energy density are normally based on the Bjorken prescription \cite{Bjorken} applied to particles emitted near mid-rapidity that probe the highest energy density of the fireball. It is, however, also of interest to use the same prescription to study how the energy density falls off away from the central region. Since the longitudinal position $z$ of a particle at the time $\tau_0$ is related to its rapidity $y$ by $z=\tau_0\sinh y$ one can study the initial energy density felt by particles as a function of rapidity, and therefore, pseudorapidity. Starting from the normal Bjorken estimate
\begin{equation}
\epsilon_0 = \frac{m_T}{\tau_0\cal{A}} \frac{dN}{dy},
\end{equation}
where $\cal{A}$ is the overlap area of the two nuclei, one may write
\begin{eqnarray}
\epsilon_0(\eta) &=& \frac{1}{\tau_0\cal{A}}\frac{dN}{d\eta} \sum_{i=\pi,K,p} P_i\\ 
\nonumber &&\int_0^\infty \frac{\sqrt{m_i^2+p_T^2}\sqrt{m_i^2+p_T^2\cosh^2 \eta}}{p_T \cosh \eta}\frac{dN}{dp_T} dp_T,
\end{eqnarray}
by transforming into $\eta$-space.

\begin{figure}[hbt]
\epsfig{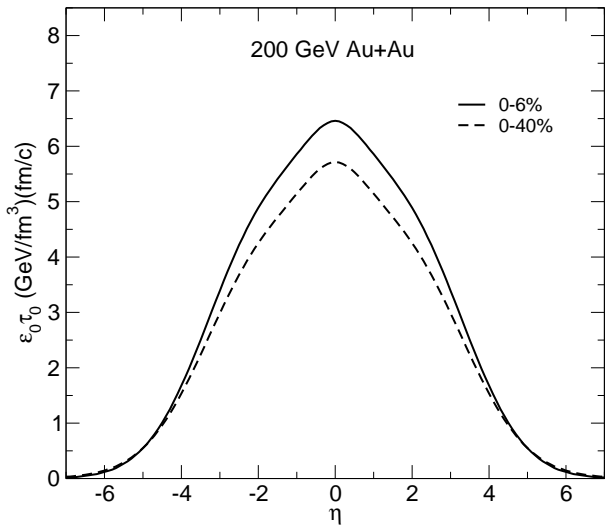}
\caption{Energy density, $\epsilon_0$ times the equilibration time, $\tau_0$ calculated for 200 GeV Au+Au collisions at 0-6\% (solid curve) and 0-40\% (dashed curve) centrality, the latter corresponding to the centrality range of the elliptical flow measurements of Ref.~\cite{PHOBOS_flow}, is shown as a function of pseudorapidity. A Gaussian $\langle p_T \rangle$ dependence on $\eta$ was used which satisfies energy conservation.}
\label{fig6}
\end{figure}

The result is illustrated in Fig. 7 for 200 GeV Au+Au collisions at 0-6\% (solid curve) and 0-40\% (dashed curve) centrality. The quantity $\epsilon_0\tau_0$ is seen to be sharply peaked at mid-rapidity and falling off rapidly for larger values of $|\eta|$. Note that this result is about 25\% larger than obtained in other works. The reason is that the overlap area, $\cal A$, for 0-6\% is actually $\sim$25 \% smaller than for a head-on collision, which has often been used. As a similar behavior has been seen in the pseudo-rapidity dependence of the elliptical flow signal, $v_2$, \cite{PHOBOS_flow} measured for 0-40\% centrality. It is reasonable to assume that the ideal liquid behavior observed at mid-rapidity via the strong flow signal is associated with the high energy density in this region. Consequently, it is tempting to speculate that there is a direct relation between the $\eta$-dependence of $v_2$ and that of the energy density $\epsilon_0$. In this context it is noteworthy that purely hydrodynamical models~\cite{Kolb} have had difficulty reproducing the observed pseudorapidity dependence of $v_2$, although recent work~\cite{buda-lund}  appears to have been more successful.

\section{Conclusions}

By requiring energy conservation in ultra-relativistic heavy-ion collisions it has been shown that unique information about the average transverse momentum of charged particles emitted in such collisions can be obtained in regions of phase space, where direct experimental measurements are not available, and, indeed, are virtually impossible. Based on $\langle p_T \rangle$ measurements over a limited rapidity region $y=0$ to $y=3.4$ by the BRAHMS collaboration, it is believed that a Gaussian $\eta$ dependence is quite accurate. Under this assumption, predictions for $\langle p_T \rangle$ are given for $Au+Au$ collisions at $\sqrt{s_{\it NN}}$=130 and 200 GeV as a function of centrality. In a recent analysis of the net-proton abundance away from mid-rapidity by the BRAHMS collaboration~\cite{BRAHMS_mesons} it is concluded that only about 73\% of the available collision energy leads to particle production, with the remaining energy being carried away by net baryons leaving the collision with near beam rapidity. Based on predictions of the HIJING/B$\overline{\rm B}$ model the present analysis shows that these particles are included in the measured $dN_{\it ch}/d\eta$-distributions. Thus, the full energy of the participating nucleons is needed to account for the observed charged-particle multiplicity. Based on the Bjorken prescription, the pseudorapidity dependence of the energy density has been derived assuming an $\eta$-independent equilibration time, $\tau_0$. For 200 GeV central (0-6\%) Au+Au collisions the mid-rapidity energy density peaks at $\sim6.5/\tau_0$ (GeV/fm$^3$), but falls off rapidly away from midrapidity. It is suggested that this mid-rapidity peaking of the energy density may be related to the observed elliptical flow signal, which exhibit a similar pseudorapidity dependence.

\section{Acknowledgements}
This work was supported by the U.S.Department of Energy, Office of Nuclear Physics, under contract No. W-31-109-ENG-38.

\section{appendix}
It is convenient to factorize the distribution $d^2N/d\eta dp_T$ such that
\begin{equation}
\frac{d^2N_i}{d\eta dp_T} = \frac{dN_i}{d\eta}\times \frac{dN_i(\eta,p_T)}{dp_T} 
\end{equation}
Furthermore it is assumed that the relative abundance of pions, kaons, and protons/anti-protons is fixed as a function of pseudo-rapidity, {\em i.e.}
\begin{equation}
\frac{dN_i}{d\eta}=P_i \frac{dN}{d\eta}.
\end{equation}
Under this assumption we may write
\begin{equation}
\frac{dE}{d\eta}=\frac{dN}{d\eta} \sum_{i=\pi,K,p}P_i \int_0^\infty \sqrt{m_i^2+p_T^2 \cosh^2\eta} \frac{dN_i}{dp_T} dp_T
\label{eq8}
\end{equation}
which is well supported by recent measurements of the identified particle distributions over a wide range of rapidity~\cite{BRAHMS_protons,BRAHMS_mesons}. As illustrated in Fig. 7a, it is, however, clear that dependence on the rest mass of the particle is important only in the mid-rapidity region, which contributes only a relatively little to the total energy.

\begin{figure}[hbt]
\epsfig{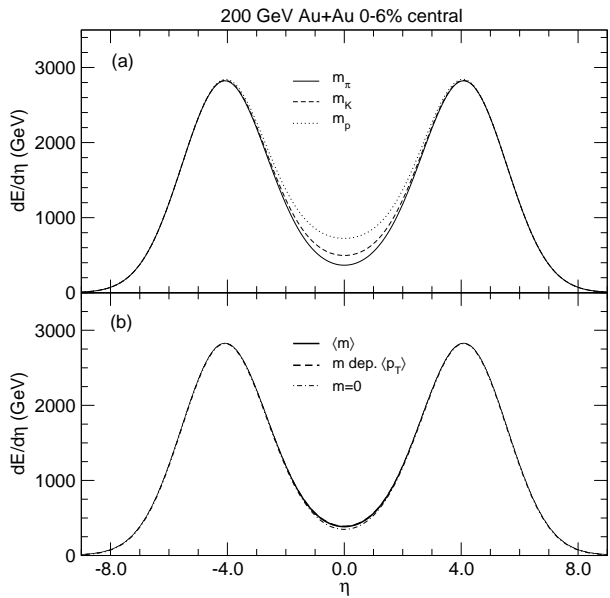}
\caption{Panel a): Comparison of $dE/d\eta$ calculated using a actual particle masses and the corresponding values of $\langle p_T \rangle_i$ (dashed curve) with one that uses an average particle mass (solid curve). A calculation with $m=0$ (dash-dotted curve) gives slightly smaller values near $\eta=0$ only. Panel b): Only in the midrapidity does the particle mass affect the energy; for $|\eta|>3$ the particle mass is irrelevant.}
\label{fig6}
\end{figure}

The present analysis is based on the expression given in Eq. 6, which furthermore assumes that the average of $dE/d\eta$ distributions is well approximated by a single distribution substituting the average particle mass, $\langle m \rangle$ obtained from the relative abundance of pions, kaons, and protons/anti-protons. The fact that this is an accurate approximation is demonstrated in Fig. 7b, where the solid curve was obtained under this assumption. It is seen to coincide with the dashed curve representing a calculation using the expression of Eq. 5 with individual values of the particle mass and mean transverse momentum of
$\langle p_T \rangle_\pi$=0.488 GeV/c, $\langle p_T \rangle_K$=0.777 GeV/c, and $\langle p_T \rangle_{p/\overline p}$=1.126 GeV/c, which were obtained from converting the values ~\cite{PHENIX_PRC69} measured in rapidity space to pseudo-rapidity. The success of this approximation is caused partly because of the relatively small abundance of kaons and protons, which is however known experimentally in the mid-rapidity region, where the total particle energy is dependent on the rest mass. At larger pseudorapidities, which account for the bulk of the energy, the results are insensitive to the particle masses.

\end{document}